\documentclass[oupdraft]{none}
\usepackage[a4paper]{geometry}
\usepackage{graphicx}
\usepackage{amsmath}
\usepackage{natbib}
\usepackage{amssymb}
\usepackage{amsfonts}
\usepackage{latexsym}
\usepackage{newlfont}
\usepackage{mathrsfs}
\usepackage{verbatim}
\usepackage{float}

\usepackage{subcaption}
\usepackage{booktabs}
\usepackage{url}
\usepackage{multirow}
\usepackage{bigstrut}
\usepackage[table,xcdraw]{xcolor}
\usepackage{pdflscape}
\usepackage{rotating}
\usepackage{cancel}
\usepackage{mathrsfs}%
\usepackage{wrapfig}

\begin{document}

\title{Modelling physical activity profiles in COPD patients: a fully functional approach to variable domain functional regression models}

\author{Pavel Hernández-Amaro$^{1\ast}$, Mar\'ia Durb\'an $^1$, M. Carmen Aguilera-Morillo$^2$,\\
Cristobal Esteban Gonzalez$^3$, Inmaculada Arostegui$^4$\\[4pt]
\textit{$^{1}$Universidad Carlos III de Madrid, \, $^{2}$Universitat Polit\`ecnica de Val\`encia, $^{3}$Osakidetza Basque Health Service, $^{4}$University of the Basque Country UPV/EHU}
\\[2pt]
E-mail address for correspondence:
{pahernan@est-econ.uc3m.es}}

\markboth%
{Hern\'andez-Amaro P. et al}
{Fully functional variable domain regression}

\maketitle

\footnotetext{To whom correspondence should be addressed.}

\begin{abstract}
{Physical activity plays a significant role in the well-being of individuals with Chronic obstructive Pulmonary Disease (COPD). Specifically, it has been directly associated with changes in hospitalization rates for these patients. However, previous investigations have primarily been conducted in a cross-sectional or longitudinal manner and have not considered a continuous perspective.
Using the telEPOC program we use telemonitoring data to analyze the impact of physical activity adopting a functional data approach. However, Traditional functional data methods, including functional regression models, typically assume a consistent data domain. However, the data in the telEPOC program exhibits variable domains, presenting a challenge since the majority of functional data methods, are based on the fact that data are observed in the same domain.
To address this challenge, we introduce a novel fully functional methodology tailored to variable domain functional data, eliminating the need for data alignment, which can be computationally taxing. Although models designed for variable domain data are relatively scarce and may have inherent limitations in their estimation methods, our approach circumvents these issues.
We substantiate the effectiveness of our methodology through a simulation study, comparing our results with those obtained using established methodologies. Finally, we apply our methodology to analyze the impact of physical activity in COPD patients using the telEPOC program's data. Software for our method is available in the form of R code on request at \url{https://github.com/Pavel-Hernadez-Amaro/V.D.F.R.M-new-estimation-approach.git}.}{B-splines; COPD;  Mixed models; Variable domain functional data}

\end{abstract}

\section{Introduction}
\label{1}

Functional data are usually found as discrete and often noisy observations of the true underlying function, measured at different locations in time, space, or other continuums. The domain where the data is observed is usually assumed to be the same across observations. Functional data where the domain is not the same for all the observations is named variable domain functional data. This type of data can be found in many data sets and a variety of research fields like biology \citep{Kulbaba2017InflorescenceEffects}, agriculture \citep{Panayi2017StatisticalRegression},  medicine \citep{Gaynanova2022ModelingSleep}, among others.

\begin{figure}[!htbp]
		\centering
		\includegraphics[scale=0.4, keepaspectratio=true]{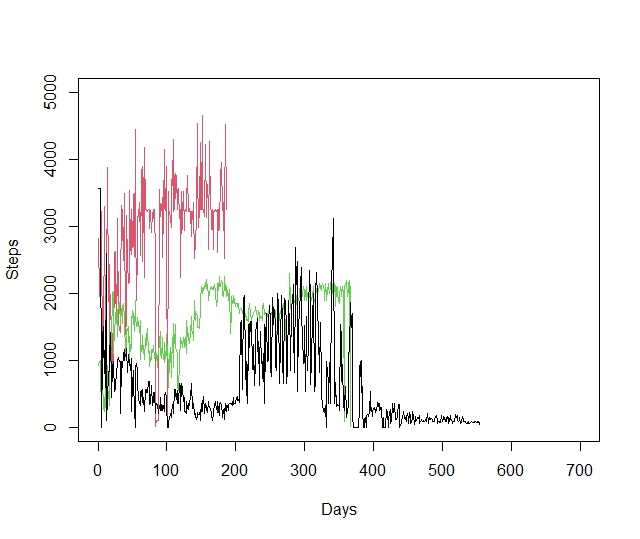}
		\caption{Daily Steps of 3 different patients of the telEPOC program.}
		\label{Pasos} 
\end{figure}

Our particular motivation is the telEPOC program \citep{Esteban2016OutcomesPatients}. In this study, a wide range of data from 119 patients suffering from Chronic Obstructive Pulmonary Disease (COPD) is collected, the most important being the physical activity performed by each patient. The physical activity is measured as daily steps, with the particularity that the number of days where steps are collected is different from patient to patient, varying from 64 days up to 1287 days, as shown in Figure \ref{Pasos}. There are two major reasons for this: (i) the exact day of sign-up in the study is different from patient to patient (the inclusion dates go from 31-05-2010 to 07-12-2013) and (ii) the time each patient spends in the study is different, being one of the causes that 21 patients died during the study period. Therefore, we are in presence of a variable domain functional data set.

One of the goals of the telEPOC program is to determine how physical activity affects health in COPD patients. More precisely, the aim is to estimate the relationship between the number of hospitalizations due to COPD (scalar discrete r.v.) and physical activity (functional covariate).

There are some studies that showed a relationship between physical activity and hospitalization for exacerbation of COPD \citep{GarciaAymerich2006RegularStudy,GarciaRio2012PrognosticCOPD}. Moreover, the change in the level of physical activity was linked to the change in hospitalizations in COPD patients \citep{Esteban2014InfluenceDisease}. These studies were cross sectional or longitudinal, observational or epidemiological. Physical activity was measured by questionnaires or by accelerometers and the tendency of the results was the same in all of them. The measurements were carried out at baseline or from time to time (for example yearly), hence, several events could have happened in that period, that could influenced the change of the level of physical activity. In a telemonitoring program (telEPOC) the information has to be more frequent, almost continuous because decisions depend on it. In other words, having daily information about physical activity is a key element in the early stages of the “decision making” process.

Therefore, we are posed with a regression problem where the response variable $Y$ (number of hospitalizations) is a scalar and the predictor $X$ (daily steps) is a function whose values vary over a continuous domain of varying length among subjects, i.e., $\{X_{i}(t): t \in [d_i,T_i], \quad i=1,\ldots,N.\}.$  Essentially, this means that every curve can have observation points that fall in different domains. In our case study, the data can be left-aligned and ordered without affecting the information of the results, having for all the data the same initial point $ t\in [0,T_i]$ and $T_i \leq T_{i+1} \, \forall i.$

To deal with the variable domain present in the data,we introduce what the authors call ``fully functional variable domain functional regression'' (FF-VDFR). The fully functional methodology assumes the sample paths (raw data) belong to a finite-dimensional space spanned by a basis of functions and hence a basis representation of the functional data is performed, jointly with a basis representation of the functional coefficient. The study of this model is the main goal of this work.

The main advantages of the proposed methodology are the following: it recovers the original functional nature of the data allowing to filter the inherent noise in the discrete observations of the sample curves, offering better performance in the context of sparse data; additionally, a flexible representation of the functional coefficient is considered in terms of flexible penalties and basis used.

In the next section we explain the main problem of modeling variable domain functional data and show the previous methodologies that deal with this problem. In Section \ref{3} we present our methodology: the FF-VDFR model, with the corresponding estimation procedure. In Section \ref{4} we present a simulation study to evaluate the performance of the proposed method in comparison with the methods introduced in Section \ref{2} showing that the FF-VDFR model outperforms the previous ones in all the evaluation criteria used. In Section \ref{5} we show the results of applying the proposed methodology to the telEPOC program. Finally we conclude with a discussion in Section \ref{6}.

\section{Variable domain functional regression}
\label{2}

In most of functional data problems, the functional predictor is assumed to have a common domain for all sample units. In this context, the sample information is usually given by $\{Y_i, X_i(t), C_i\}$, $i = 1, \ldots, N$, where $C_i$ is a vector of non-functional covariates, $Y_i$ is a scalar outcome following an exponential family distribution with mean $\mu_i$ and $\{X_{i} (t) : t \in T\}$ is the functional predictor. From this information, the functional generalized linear model is given by 
\begin{equation}
\label{s-o-f}
\eta_i = g(\mu_i) = \alpha + \boldsymbol{C_i\gamma} + \int_0^T X_i(t)\beta(t) \, \text{d}t,
\end{equation}
with $g(\cdot)$ being the corresponding link function. 

This model, named scalar-on-function regression model (SOF), was one of the first regression models extended to the case of functional data. The main theoretical aspects related to this model were studied in \cite{Cardot1999} and in \cite{James2002}, in the more general framework of generalized linear models. This model has been widely used in the literature, leading to numerous applications and new methodological developments. Penalized versions of the functional generalized linear models can be seen in works, such as \cite{Cardot2005EstimationLikelihood}, \cite{Goldsmith2011PenalizedRegression}, \cite{Aguilera-Morillo2013PenalizedRegression}, among others. 

This methodology is very useful for modeling functional data when the domain is constant across observations, but to deal with variable domain functional data a previous transformation of the sample curves is needed. One of the most common choice is based on the registration of the sample curves to a common domain. This additional step presents some drawbacks in the case of variable domain functional data: information given by the specific shape of the curves can be lost, adding errors in the estimation of the curves; the resulting estimation of the functional coefficient will be more difficult or impossible to interpret since it will be a single curve estimated from the forced common domain; this registration procedure could be computational expensive in some cases. For some insights in registration of curves see \cite{Ramsay2005, Ramsay2009FunctionalMATLAB}. 

Moreover, in a variable domain dataset, the length of the sample paths might be informative itself. Therefore, incorporating this information in the formulation of the functional regression model is essential for dealing with this type of data. 

As solution to the drawbacks of using the SOF model in the case of variable domain functional data \cite{Gellar2014Variable-DomainData} proposed a new model, the variable domain regression model (VDFR): 
\begin{equation}
\label{Gellar}
\eta_i = g(\mu_i) =\alpha + \boldsymbol{C_i\gamma} + \frac{1}{T_i}\int_{0}^{T_i} X_i(t)\beta(t,T_i) \, \text{d}t,\ \ t\in [0,T_i],
\end{equation}
where the univariate coefficient function $\beta(t)$ is now replaced by the bivariate coefficient function $\beta(t, T)$, that now depends on the time instant $t$ and the data domain $T$. This functional coefficient is now a surface and the curves obtained by fixing the variable $T=T_i$ represent the optimal function for $X_i(t)$ to express its contribution over $g(\mu_i)$. Moreover, the integration limits, previously fixed to be from 0 to $T$, are now subject-specific, avoiding then the necessity of a previous curve registration. The authors use directly the discrete observations $x_{ik}$ of each sample curve $X_i(t)$ collected at a finite set of observation points ${t_{ik}: k=0,...,m_i}$, ignoring the functional nature of the curves. As a result, the methodology proposed by \cite{Gellar2014Variable-DomainData} can be categorized as "partially functional approach", since it does not take into consideration the functional form of the variable $X(t)$. Instead,  the methodology works directly with the raw data matrix, comprising sample curves at the observation points. This approach bypasses the opportunity to recover the genuine functional form of the data, leaving out the possibility of recovering the true functional form of the data and not filtering the noise commonly found in discrete observations, which subsequently introduces errors to the model coefficients.  

Additionally, this approach recommends the use of thin-plate spline functions for the basis representation of the functional coefficient, using an isotropic penalization, which forces the same degree of smoothness in both dimensions of the functional coefficient. This might result in biased estimates of the functional coefficient, as it will be shown in the simulation section.

\section{Fully functional Variable domain functional regression}
\label{3}

As we mentioned in the previous section, partially functional approaches work on the discrete observations $x_{ij}$ of each sample curve  $X_i(t)$ at a set of points $\{t_{ik}, \quad k=0,\ldots T_i\}.$ However, in practice it is very common to find functional datasets observed with error or noise. In that sense, a fully functional approach will perform a pre-smoothing of the sample curves, recovering the smooth functional form of the data by means of a basis representation of the sample curves; this is the approach taken in this section. A review on the different ways to estimate the basis coefficients as well as the different penalties used and their performance is shown in \cite{Aguilera2013ComparativeData}.

\subsection{Model formulation in terms of basis functions}
\label{3.1}
Let $Y$ be the scalar response variable and $X(t)$ the functional predictor. Let $X (t)$ be a second order continuous-time stochastic process, with sample functions $\{X_i(t): t \in [d_i,T_i], \; i=1,\ldots,N\}$ in the Hilbert space $\mathscr{H}_1 = L^2(T)$ of square integrable functions, with the usual inner product:
$$<f,g> = \left\{f: T\leftarrow \mathbb{R}: \int_T f^2(t) \, \text{d}t < \infty \right\}.$$

Let us assume the basis representation of the sample curves and the functional coefficient of model (\ref{Gellar}) as follows:
\begin{eqnarray*}
 	X_i(t) &=& \displaystyle \sum_{j=1}^{p_i}a_{ij}\phi_{j}(t)=\boldsymbol{\phi}_i'(t)\boldsymbol{a}_i,\\
	\beta(t,T)  &=&  \sum_{l=1}^{q}\sum_{k=1}^{r}b_{lk }\varphi_{l}(t)\psi_{k}(T) = \boldsymbol{M}(t,T)\boldsymbol{b},
\end{eqnarray*}
where $ \boldsymbol{\phi}_i(t) = (\phi_{1}(t), \phi_{2}(t), \ldots, \phi_{p_i}(t))'$ and $\boldsymbol{M}(t,T)$ are the basis used in the representation of the functional data and the functional coefficient and $\boldsymbol{a}_i$ and $\boldsymbol{b}$ their respective basis coefficients. Notice that $\boldsymbol{M}(t,T)$ is a bivariate basis function resulting from the tensor product of $\boldsymbol{\varphi}(t)$ and $\boldsymbol{\psi}(T),$ with $ \boldsymbol{\psi}(t) = (\psi_1(t), \psi_2(t), \ldots, \psi_{r}(t))' $ and $ \boldsymbol{\varphi}(t) = (\varphi_1(t), \varphi_2(t), \ldots, \varphi_{q}(t))'.$

Here $p_i, q$ and $r$ are the respective number of basis of $\boldsymbol{\phi}_i(t), \boldsymbol{\varphi}(t)$ and $\boldsymbol{\psi}(t)$. For simplicity, hereinafter the same number of basis ($p$) is considered for the basis representations of all sample curves, i.e., $p_i=p$ and $\boldsymbol{\phi}_i(t) = \boldsymbol{\phi}(t) \; \forall i = 1, \ldots, N$, but this can be easily relaxed.

The choice of the basis is important. This decision is often data driven: if data have periodic trends a Fourier basis can be used; if data present a strong locally behavior and its derivatives are not of interest, wavelets basis are the common choice. In this paper B-splines basis \citep{DeBoor2001ASplines} have been considered, which is the common choice when the underlying signal is assumed to be smooth and their derivatives up to a certain order are needed.  

As a consequence of assuming the basis representation of both, sample curves and functional coefficient, model (\ref{Gellar}) turns into the following multivariate regression model.

\begin{equation}		
 \boldsymbol{\eta} =\boldsymbol{\alpha} + \boldsymbol{C}\boldsymbol{\gamma} + \frac{1}{\boldsymbol{T}}\int_0^{\boldsymbol{T}} X(t)\beta(t,T)\; \text{d}t =\boldsymbol{\alpha} + \boldsymbol{C}\boldsymbol{\gamma} +\boldsymbol{A}\boldsymbol{\Psi}\boldsymbol{b}=\boldsymbol{B}\boldsymbol{\theta},
\end{equation}
\label{modelo multivariante}
where $\boldsymbol{T}$ represents the vector considering the length of all curves, so for each sample curve the integration limits are different. 

The matrix of coefficients $\boldsymbol{A}$ is a block diagonal matrix, where the $i$-th block of the diagonal is the estimated vector of basis coefficients $\boldsymbol{a}_i'$. To estimate these basis coefficients, we follow \cite{Aguilera2013ComparativeData}; specifically, penalized least squares regression is used, incorporating a discrete penalty term that relies on the second-order differences between neighboring coefficients of the B-splines \citep{Eilers1996FlexiblePenalties}. Finally the matrix of inner products $\boldsymbol{\Psi}$ is a block column matrix of weighted inner products, with the $i$-th block being a weighted inner product between the basis $\boldsymbol{\phi}(t)$ and $\boldsymbol{M}(t,T_i)$, i.e., $\boldsymbol{\Psi}_{Np\times qr} = (\boldsymbol{\Psi}_1, \ldots, \boldsymbol{\Psi}_N)'$, where $\boldsymbol{\Psi}_i = \frac{1}{T_i}\langle \boldsymbol{\phi}(t),\boldsymbol{M}(t,T_i)\rangle$.
\begin{eqnarray*}
     \boldsymbol{A} =  \begin{pmatrix}
		\boldsymbol{a}_1' & 0 & \ldots & 0\\
		0 & \boldsymbol{a}_2' & 0 & \ldots\\
		\ldots & \ldots & \ldots & \ldots\\
		0 & 0 & \ldots & \boldsymbol{a}_N' 	
	\end{pmatrix}_{N\times Np} &    	    \boldsymbol{\Psi} =  \begin{pmatrix}
	\vspace{0.1cm}\frac{1}{\boldsymbol{T}_1}\int_0^{T_1}\boldsymbol{\phi}(t)\boldsymbol{M}(t,T_1)\; \text{d}t \\
	\frac{1}{\boldsymbol{T}_2}\int_0^{T_2}\boldsymbol{\phi}(t)\boldsymbol{M}(t,T_2)\; \text{d}t \\
	\vdots\\
	\frac{1}{\boldsymbol{T}_N}\int_0^{T_N}\boldsymbol{\phi}(t)\boldsymbol{M}(t,T_N)\; \text{d}t
\end{pmatrix}_{Np\times qr}
	\end{eqnarray*}

The basis representation of the curves adds some difficulty since it is necessary to calculate the inner product between a univariate and bivariate function. This has been recently solved by \cite{Masak2022SeparableProduct} with the following result: 

\textbf{Proposition 1:} Let $\mathscr{H}_1=L^2(T)$ and $\mathscr{H}_2=L^2(F)$ be two separable Hilbert spaces as in Section \ref{1} with $F=\{T: T_{min} \leq T \leq T_{max}\}$ being the space corresponding to all the different values of the data domains.

Let $\mathscr{H} = \mathscr{H}_1 \otimes \mathscr{H}_2$ where $\otimes$ represents the tensor product and let $f(t), \, u(T)$ and $h(t,T)$ be functions in $\mathscr{H}_1, \, \mathscr{H}_2$ and $\mathscr{H},$ respectively
 
Then the partial inner products are two unique bi-linear operators $K_1 : \mathscr{H} \times \mathscr{H}_1  \rightarrow \mathscr{H}_2 $ and $K_2 : \mathscr{H} \times \mathscr{H}_2  \rightarrow \mathscr{H}_1 $ defined by:
\begin{eqnarray*}
    K_1(T)_{h, f} &=& \int_{T}f(t) h(t,T) \; \text{d}t\\
    K_2(t)_{h, u} &=& \int_{F}u(T) h(t,T) \; \text{d}T \centerdot
\end{eqnarray*}

Using this result, the elements of the new matrix of inner products are given by $\boldsymbol{\Psi}_{Np\times qr} = (\boldsymbol{\Psi}_1, \ldots, \boldsymbol{\Psi}_N)'$ are $ \boldsymbol{\Psi}_i = \displaystyle \frac{1}{T_i}K_1(T_i)_{M, \phi} = \displaystyle \frac{1}{T_i}\int_{T_i}\boldsymbol{\phi}(t)\boldsymbol{M}(t,T_i) \; \text{d}t$.

These integrals are numerically approximated by the Composite Simpson method. The key difficulty is to perform the integration only in the $t$ dimension while maintaining the proper two dimensional structure of the basis $M(t,T_i)$. In practice this problem is overcome by the following, for each iteration of the integration, a matrix $\boldsymbol{M}_i$ is obtained by performing the Kronecker product of two matrices $\boldsymbol{\varphi}_i$ and $\boldsymbol{\psi}_i$. The matrix $\boldsymbol{\psi}_i$ is the result of evaluating the basis $\boldsymbol{\psi}(T)$ in the corresponding domain $T_i$, i.e., $\boldsymbol{\psi}_i = \boldsymbol{\psi}(T_i)$, and matrix $\boldsymbol{\varphi}_i$ is the result of evaluate the basis $\boldsymbol{\varphi}(t)$ in a set of points determined by the integration method (this set of points change according with the domain of every curve). The basis $\boldsymbol{\phi}(t)$ is evaluated in the same set of points as the basis $\boldsymbol{\varphi}(t)$ in every iteration resulting in a matrix $\boldsymbol{\phi}_i$. Finally when the matrix $\boldsymbol{M}_i$ is recalculated, the product between $\boldsymbol{M}_i$ and $\boldsymbol{\phi}_i$ is performed.

Notice that the matrix $\boldsymbol{\psi}_i$ is the $i$-th row of a more general matrix $(\boldsymbol{\psi})_{N\times r}$, associated to all the different domains present in the data: $\boldsymbol{T}=[T_1, \ldots, T_N]$ and then, for subject $i$, we select the corresponding row. Two or more different sample curves can have the same domain; in this case the corresponding row of the matrix $\boldsymbol{\psi}$ will be the same for all of them.

We use B-splines for all our basis representations because of their desirable properties, but this is not a restriction.

\subsection{Model estimation through a mixed model representation}
\label{3.2}

The multivariate regression model (3.3) falls into the category of generalized linear models and therefore the maximum likelihood method is used in order to estimate the model parameters. In our motivational example the response variable follows a Poisson distribution with likelihood: 

$$L(\boldsymbol{\theta},\boldsymbol{y}) = \displaystyle\sum_{i=1}^N y_i \eta_i - exp\left\{ \sum_{i=1}^N \eta_i\right\}.$$

Since the functional coefficient has been represented using a B-spline basis, the smoothness of the resulting estimated coefficient is determined by the basis dimension. To avoid the problem of choosing the optimal number of basis functions we follow the penalized likelihood approach by \cite{Eilers1996FlexiblePenalties} with the final penalized likelihood equation: 

$$L_p(\boldsymbol{\theta},\boldsymbol{y}) = L(\boldsymbol{\theta},\boldsymbol{y}) - \frac{1}{2} \boldsymbol{\theta' P\theta},$$

where $L(\boldsymbol{\theta},\boldsymbol{y})$ is the likelihood of $\boldsymbol{Y}$ and $ \boldsymbol{P}$ is the penalty term. Penalties are, in general, based on derivatives of curves \citep{Wood2017P-splinesData} or differences between adjacent B-splines coefficients \citep{Eilers1996FlexiblePenalties}. We take here this second approach.

Considering that the functional parameter is two dimensional, an anisotropic two dimensional penalization is used, allowing to control the smoothness of the functional coefficient independently for each dimension. The penalization added is:
\begin{equation}
     \boldsymbol{P} = \lambda_t(\boldsymbol{I}_r \times \boldsymbol{D}_t' \boldsymbol{D}_t')+\lambda_T(\boldsymbol{I}_q \times \boldsymbol{D}_T' \boldsymbol{D}_T),
\end{equation}

where the matrices $\boldsymbol{D}_t$ and $\boldsymbol{D}_T$ are second order differences matrices, with $\times$ denoting the Kronecker product \citep{lee2011p}.

This penalized approach makes the choice of the number of basis not relevant (provided that the size of the basis is large enough), controlling the smoothness through the smoothing parameters $\lambda_t$ and $\lambda_T$.
 
Finally, we use the mixed model reparametrization of a penalized spline to estimate the parameters of the FF-VDFR model. This transformation allows the estimation of all parameters in the model, including the smoothing parameters, simultaneously. A brief description of this reparametrization is done next to help the reader understand the used methodology. For a more detailed insight into the mixed model reparametrization of a penalized spline when functional data do not present variable domain see \cite{Lee2010SmoothingData}. Our aim is to transform
\begin{equation}
    \boldsymbol{\eta}=\boldsymbol{B}\boldsymbol{\theta} \Rightarrow \boldsymbol{X}\boldsymbol{\nu} + \boldsymbol{Z}\boldsymbol{\delta},\quad \boldsymbol{\delta}\sim N(0,\boldsymbol{G}),
\end{equation}\label{mixed model}
with $\boldsymbol{X}$ and $\boldsymbol{Z}$ being the model matrices, $\boldsymbol{\nu}$ and $\boldsymbol{\delta}$ the fixed and random effects respectively, and $\boldsymbol{G}$ the variance-covariance matrix of the random effects, which depends on two variance components $\tau_t^2$ and $\tau_T^2$. This reparametrization is done through a transformation matrix $ \boldsymbol{T} $ based on the Singular Value Decomposition (SVD) factorization of the product of the differences matrices $ \boldsymbol{D}_i' \boldsymbol{D}_i$. Let
	\begin{equation*}
	    \boldsymbol{D}_i' \boldsymbol{D}_i= \left[\boldsymbol{U}_{in}|\boldsymbol{U}_{is}\right]\left[\begin{matrix}
		\boldsymbol{0}_2 & \\
		& \tilde{\boldsymbol{\Sigma}}_i
	\end{matrix}\right]\left[\begin{matrix}
		\boldsymbol{U}_{in}'\\
		\boldsymbol{U}_{is}'
	\end{matrix}\right]
	\end{equation*}  

be the SVD factorization of the matrix $ \boldsymbol{D}_i' \boldsymbol{D}_i$, for $i=\{t,T\}$, where $\boldsymbol{U}_{in}$ and $\boldsymbol{U}_{is}$ are the eigenvectors associated with the zero and non-zero eigenvalues, respectively. Then the transformation matrix $ \boldsymbol{T} $ is define as: $
    \boldsymbol{T}=\left[\boldsymbol{T}_n|\boldsymbol{T}_s\right] = \left[\boldsymbol{U}_{Tn}\times \boldsymbol{U}_{tn}|\boldsymbol{U}_{Ts}\times \boldsymbol{U}_{tn}:\boldsymbol{U}_{Tn}\times \boldsymbol{U}_{ts}:\boldsymbol{U}_{Ts}\times \boldsymbol{U}_{ts}\right].
$

Other options for the transformation matrix $\boldsymbol{T}$ are possible, but the one proposed in this paper allows to recover the estimated original parameter $\boldsymbol{\widehat{\theta}}$ from the estimated mixed model coefficients thanks to the orthogonality property of the matrix $\boldsymbol{T}$ and, hence, recover the estimated functional coefficient $\widehat{\beta}(t,T)$. Using this transformation matrix, the model (\ref{modelo multivariante}) is reparametrized as follows: $\boldsymbol{\eta}=\boldsymbol{B}\boldsymbol{\theta} =\boldsymbol{B}\boldsymbol{T}\boldsymbol{T}'\boldsymbol{\theta} =\boldsymbol{X}\boldsymbol{\nu} + \boldsymbol{Z}\boldsymbol{\delta},$ where $ \boldsymbol{B}\boldsymbol{T}= \left[\boldsymbol{B}\boldsymbol{T_n}|\boldsymbol{B}\boldsymbol{T_s}\right]=\left[\boldsymbol{X}|\boldsymbol{Z}\right] $, $\boldsymbol{T}'\boldsymbol{\theta}=\boldsymbol{\omega}$, with $ \boldsymbol{\omega}'=(\boldsymbol{\nu}',\boldsymbol{\delta}') $ and the variance-covariance matrix $\boldsymbol{G}$ is obtained from applying this transformation to the penalization used before, $\boldsymbol{G}^{-1}= \boldsymbol{T}' \boldsymbol{P} \boldsymbol{T}$, with
\begin{equation}
\boldsymbol{G}^{-1} = \begin{pmatrix}
		\dfrac{1}{\tau_T^2}\tilde{\boldsymbol{\Sigma}}_T \times \boldsymbol{I}_{2} & & \\
		& \dfrac{1}{\tau_t^2} \boldsymbol{I}_{2} \times \tilde{\boldsymbol{\Sigma}}_t & \\
		& & \dfrac{1}{\tau_T^2}\tilde{\boldsymbol{\Sigma}}_T \times \boldsymbol{I}_{q-2} + \dfrac{1}{\tau_t^2}\boldsymbol{I}_{r-2} \times \tilde{\boldsymbol{\Sigma}}_t
	\end{pmatrix},
\end{equation}

where for the variance components we have the relations $\tau_t^2 = \dfrac{1}{\lambda_t}$ and $\tau_T^2 = \dfrac{1}{\lambda_T}$ \citep{Brumback1999VariableComment}.

Finally, penalized quasi-likelihood \citep{Breslow1993ApproximateModels} is used to estimate the mixed model coefficient. In order to speed up computations, the SOP algorithm has been applied \citep{Rodriguez-Alvarez2019OnSmoothing}.

\section{Simulation study}
\label{4}

To assess the performance of the FF-VDFR model, we conducted a simulation study in this section and compared the results obtained with those from both the VDFR model and the conventional scalar-on-function (SOF) regression models. In the case of the SOF model, a previous registration of the curves was performed. The simulation scheme is inspired by the one performed in \cite{Gellar2014Variable-DomainData}.

\subsection{Simulation scenarios}
\label{4.1}

For simplicity, only models with one functional covariate and no non-functional covariates have been considered. In this study 100 data sets have been simulated for each combination of the following parameters in a total of $3\times 2\times 2\times 2\times 4=96$ different scenarios:

\begin{itemize}
    \item Three sample sizes: $N=\{ 100, 200, 500\}$.
    \item Two different types of outcomes: continuous data and count data. In both cases the following linear predictor is used:    
\begin{equation*}
    \eta_i = \dfrac{1}{T_i} \displaystyle \sum_{t=1}^{T_i} X_i(t) \beta(t,T_i), \;\; t=1, \ldots, T_i\leq 100, 
		\end{equation*}


with $Y_i = \eta_i + \epsilon_i, \; \epsilon_i\sim N(0, 1)$ for the continuous outcome and $Y_i \sim \text{Poiss}(\mu_i)$ with $\mu_i = \text{exp}(\eta_i)$ for the count data. 

    \item Two different distribution for the data domain $T_i$: Uniform($T_i \sim U(10,100)$) and Negative Binomial ($T_i \sim NegBin(1, p= 0,04)$).
    
        The domain of the sample curves is set to ensure that every curve have a minimum of 10 and a maximum of 100 observations in both distribution settings (we considered 10 to be an acceptable minimum to consider the observed points as observation of the true underlying functional data $X_i(t)$). When $T_i$ is simulated using a negative binomial distribution we truncate the generated values to belong in the interval $(10,100)$, so that when a generated value is lower than 10 (higher than 100) this is set by default as 10 (100).

    \item Two levels of noise for the true functional covariate.
    
    The true functional covariate $X_i(t)$ is simulated according to the following:
\begin{equation*}
	X_i(t) = u_i + \displaystyle \sum_{k=1}^{10}\left\{ v_{ik1}\cdot sen\left(\frac{2\pi k}{100}t\right) + v_{ik2} \cdot \cos\left(\frac{2\pi k}{100}t\right)\right\} + \delta_i(t),
\end{equation*}

with $u_i\sim N(0,1), \; v_{ik1}, v_{ik2}\sim N(0,\frac{4}{k^2}), \; \delta_i(t)\sim N(0, \sigma_x), \; t=1, \ldots, T_i \leq 100$ and  $\sigma_x=\{0,1\}$. Here $\sigma_x=0$ indicates that the true functional data has been considered as smooth curves and $\sigma_x=1$ indicates that the true functional data has been considered as noisy curves. 

    \item Four different possibilities for the functional coefficient $\beta(t,T)$ defined by 
\end{itemize}
\begin{eqnarray*}
		& \beta_1(t,T_i)  =  10 \dfrac{t}{T_i} - 5 \,; & \beta_2(t,T_i) =  \left(1-\dfrac{2T_i}{T}\right)  \times  \left(5 - 40\left(\dfrac{t}{T_i} - 0.5\right)^2\right)\,;\\
		& \beta_3(t,T_i) =  5 - 10\left(\dfrac{T_i-t}{T}\right)\,; &\; \beta_4(t,T_i) = sen\left(\dfrac{2\pi T_i}{T}\right)  \times  \left( 5 -10\left(\dfrac{T_i - t}{T}\right)\right),
	\end{eqnarray*} 

where $T = \max\{T_1, \ldots, T_N\}=T_N$.

In both the FF-VDFR and SOF models, 25 cubic B-splines (with p=25) were employed to represent the functional data. For the functional coefficient, the SOF model utilized 25 cubic B-splines, while the FF-VDFR model incorporated 25 basis functions for both the marginal basis (with q=r=25), resulting in a bi-dimensional basis of size 625.

The penalty terms utilized for estimating the FF-VDFR and SOF models were based on a second-order differences matrix. The VDFR model considers a thin plate basis of size 89 for the functional coefficient, which is the maximum size allowed by the software when $T_N=100$, and an isotropic penalty based on second-order derivatives.\\

All simulations were implemented in \cite{RCoreTeam2013R:Computing}. The package \textbf{SOP} \citep{Rodriguez-Alvarez2021SOP:Estimation} have been used for the mixed model reparametrization of the multivariate regression model and the estimation of all model parameters. The estimation of the SOF and VDFR models have been performed using the \textbf{refund} package \citep{Goldsmith2021Refund:Data}.

\subsection{Performance criteria}

We evaluate the performance of the models with respect to two important aspects. The first one is the prediction ability. To this end, a cross-validation 10-fold approach has been carried out. The measure that we use for this prediction errors is the mean of the root mean square error (RMSE) calculated for every fold: 
\begin{equation*}
	RMSE_j = \sqrt{\frac{\displaystyle \sum_{i=1}^{N_j} (Y_{ij} - \hat{Y}_{ij})^2}{N_j}}, \; \; j=1, \ldots, 10,
\end{equation*}

where $Y_{ij}$ is the $i$-th response variable in the $j$-th fold, $\hat{Y}_{ij}$ its corresponding estimation and $N_j$ is the number of responses in the $j$-th fold.

Another important aspect is the capability to accurately estimate the functional coefficient $\beta(t,T)$. In this case we use  the Average Mean Square Error (AMSE) defined as follows:

\begin{equation*}
	AMSE^r = \frac{1}{T(T+1)} \displaystyle \sum_{k=10}^{T} \sum_{t=1}^{k} \left\{\beta(t,k)-\hat{\beta}(t,k)\right\}^2,   
\end{equation*}

where $T = \max\{T_1, \ldots, T_N\} = T_N$ and $\hat{\beta}(t,k)$ is the estimated functional coefficient.

Notice that with the SOF model it is not possible to calculate the AMSE since this model only considers one fixed domain for the functional coefficient.

\subsection{Results}

In this section we comment the results obtained in the simulation study for all the scenarios, but due to lack of space only tables and figures for the scenarios where $T_i$ is generated from the negative binomial distribution and the response variable was simulated from a Poisson distribution are shown. This scenario was chosen because it best reflects the nature of the data from the telEPOC program. The rest of the tables and figures can be found in the supplementary material.

Table \ref{Y-tables} and Table \ref{B-tables} show the mean and standard deviation (in parenthesis) of the RMSE and AMSE, respectively, for all the possible true coefficient functions and when the true functional data are smooth or noisy. The lowest values are highlighted. 

\begin{table}[htbp]
\resizebox{\columnwidth}{!}{%
  \centering
    \begin{tabular}{|c|c|c|c|c|c|c|c|c|}
\cline{2-9}    \multicolumn{1}{c|}{\multirow{3}[6]{*}{}} & \multicolumn{8}{c|}{N=100} \bigstrut\\
\cline{2-9}    \multicolumn{1}{c|}{} & \multicolumn{4}{c|}{Smooth data} & \multicolumn{4}{c|}{Noisy data} \bigstrut\\
\cline{2-9}    \multicolumn{1}{c|}{} & $\beta_1$ & $\beta_2$ & $\beta_3$ & $\beta_4$ & $\beta_1$ & $\beta_2$ & $\beta_3$ & $\beta_4$ \bigstrut\\
    \hline
    VDFR  & 1.343 (0.256) & 1.145 (0.141) & 3.143 (1.331) & 1.924 (0.532) & 1.364 (0.279) & 1.155 (0.147) & 2.793 (1.047) & 1.991 (0.617) \bigstrut\\
    \hline
    SOF   & 5.350 (3.762) & 1.813 (0.851) & 3.196 (0.867) & 2.798 (1.147) & 5.128 (3.039) & 1.803 (0.979) & 5.199 (2.471) & 2.543 (0.881) \bigstrut\\
    \hline
    FF-VDFR & \textbf{1.135 (0.134)} & \textbf{1.118 (0.12)} & \textbf{2.423 (0.898)} & \textbf{1.862 (0.568)} & \textbf{1.137 (0.136)} & \textbf{1.148 (0.142)} & \textbf{2.121 (0.667)} & \textbf{1.824 (0.553)} \bigstrut\\
    \hline
    \multicolumn{1}{c|}{\multirow{3}[6]{*}{}} & \multicolumn{8}{c|}{N=200} \bigstrut\\
\cline{2-9}    \multicolumn{1}{c|}{} & \multicolumn{4}{c|}{Smooth data} & \multicolumn{4}{c|}{Noisy data} \bigstrut\\
\cline{2-9}    \multicolumn{1}{c|}{} & $\beta_1$ & $\beta_2$ & $\beta_3$ & $\beta_4$ & $\beta_1$ & $\beta_2$ & $\beta_3$ & $\beta_4$ \bigstrut\\
    \hline
    VDFR  & 1.193 (0.104) & 1.10 (0.076) & 2.693 (0.59) & 1.827 (0.341) & 1.206 (0.117) & 1.102 (0.077) & 2.633 (0.768) & 1.831 (0.38) \bigstrut\\
    \hline
    SOF   & 5.269 (2.865) & 2.316 (1.146) & 4.153 (0.88) & 2.923 (0.828) & 5.780 (3.422) & 2.279 (1.123) & 5.641 (2.315) & 2.953 (0.974) \bigstrut\\
    \hline
    FF-VDFR & \textbf{1.120 (0.101)} & \textbf{1.092 (0.077)} & \textbf{2.311 (0.607)} & \textbf{1.690 (0.343)} & \textbf{1.111 (0.1)} & \textbf{1.096 (0.084)} & \textbf{1.858 (0.33)} & \textbf{1.699 (0.39)} \bigstrut\\
    \hline
    \multicolumn{1}{c|}{\multirow{3}[6]{*}{}} & \multicolumn{8}{c|}{N=500} \bigstrut\\
\cline{2-9}    \multicolumn{1}{c|}{} & \multicolumn{4}{c|}{Smooth data} & \multicolumn{4}{c|}{Noisy data} \bigstrut\\
\cline{2-9}    \multicolumn{1}{c|}{} & $\beta_1$ & $\beta_2$ & $\beta_3$ & $\beta_4$ & $\beta_1$ & $\beta_2$ & $\beta_3$ & $\beta_4$ \bigstrut\\
    \hline
    VDFR  & 1.129 (0.053) & 1.087 (0.053) & 2.721 (0.66) & 1.675 (0.216) & 1.133 (0.053) & \textbf{1.090 (0.053)} & 2.330 (0.452) & \textbf{1.692 (0.243)} \bigstrut\\
    \hline
    SOF   & 3.569 (0.85) & 2.616 (1.089) & 3.848 (0.705) & 3.035 (0.848) & 3.523 (0.831) & 2.487 (1.067) & 9.386 (1.221) & 3.143 (0.893) \bigstrut\\
    \hline
    FF-VDFR & \textbf{1.101 (0.056)} & \textbf{1.084 (0.056)} & \textbf{2.374 (0.508)} & \textbf{1.65 (0.288)} & \textbf{1.096 (0.05)} & 1.091 (0.061) & \textbf{1.964 (0.255)} & 1.695 (0.323) \bigstrut\\
    \hline
    \end{tabular}}%
\caption{Mean (standard deviation) of 100 measures of RMSE for all the scenarios where the domain follows a negative binomial distribution and the response follows a Poisson distribution.}
  \label{Y-tables}%
\end{table}%

\begin{table}[htbp]
\resizebox{\columnwidth}{!}{%
  \centering
    \begin{tabular}{|c|c|c|c|c|c|c|c|c|}
\cline{2-9}    \multicolumn{1}{c|}{\multirow{3}[6]{*}{}} & \multicolumn{8}{c|}{N=100} \bigstrut\\
\cline{2-9}    \multicolumn{1}{c|}{} & \multicolumn{4}{c|}{Smooth data} & \multicolumn{4}{c|}{Noisy data} \bigstrut\\
\cline{2-9}    \multicolumn{1}{c|}{} & \multicolumn{1}{l|}{$\beta_1$} & \multicolumn{1}{l|}{$\beta_2$} & \multicolumn{1}{l|}{$\beta_3$} & \multicolumn{1}{l|}{$\beta_4$} & \multicolumn{1}{l|}{$\beta_1$} & \multicolumn{1}{l|}{$\beta_2$} & \multicolumn{1}{l|}{$\beta_3$} & \multicolumn{1}{l|}{$\beta_4$} \bigstrut\\
    \hline
    VDFR  & 0,0187 (0,0086) & \textbf{0,0155 (0,0052)} & 0,0118 (0,0068) & 0,0196 (0,013) & 0,0197 (0,009) & \textbf{0,0163 (0,0054)} & 0,0118 (0,0053) & \textbf{0,0215 (0,0126)} \bigstrut\\
    \hline
    FF-VDFR & \textbf{0,0079 (0,0075)} & 0,0176 (0,0074) & \textbf{0,008 (0,0126)} & \textbf{0,018 (0,0199)} & \textbf{0,008 (0,0073)} & 0,0177 (0,0067) & \textbf{0,0083 (0,0103)} & 0,0229 (0,0231) \bigstrut\\
    \hline
    \multicolumn{1}{c|}{\multirow{3}[6]{*}{}} & \multicolumn{8}{c|}{N=200} \bigstrut\\
\cline{2-9}    \multicolumn{1}{c|}{} & \multicolumn{4}{c|}{Smooth data} & \multicolumn{4}{c|}{Noisy data} \bigstrut\\
\cline{2-9}    \multicolumn{1}{c|}{} & \multicolumn{1}{l|}{$\beta_1$} & \multicolumn{1}{l|}{$\beta_2$} & \multicolumn{1}{l|}{$\beta_3$} & \multicolumn{1}{l|}{$\beta_4$} & \multicolumn{1}{l|}{$\beta_1$} & \multicolumn{1}{l|}{$\beta_2$} & \multicolumn{1}{l|}{$\beta_3$} & \multicolumn{1}{l|}{$\beta_4$} \bigstrut\\
    \hline
    VDFR  & 0,0246 (0,0039) & 0,0217 (0,0057) & 0,0142 (0,0051) & 0,0254 (0,0101) & 0,0243 (0,0044) & 0,022 (0,0063) & 0,013 (0,0043) & 0,0264 (0,0107) \bigstrut\\
    \hline
    FF-VDFR & \textbf{0,0083 (0,0094)} & \textbf{0,0193 (0,0098)} & \textbf{0,0085 (0,0058)} & \textbf{0,021 (0,0084)} & \textbf{0,0096 (0,0088)} & \textbf{0,019 (0,0093)} & \textbf{0,0055 (0,0053)} & \textbf{0,0208 (0,0076)} \bigstrut\\
    \hline
    \multicolumn{1}{c|}{\multirow{3}[6]{*}{}} & \multicolumn{8}{c|}{N=500} \bigstrut\\
\cline{2-9}    \multicolumn{1}{c|}{} & \multicolumn{4}{c|}{Smooth data} & \multicolumn{4}{c|}{Noisy data} \bigstrut\\
\cline{2-9}    \multicolumn{1}{c|}{} & \multicolumn{1}{l|}{$\beta_1$} & \multicolumn{1}{l|}{$\beta_2$} & \multicolumn{1}{l|}{$\beta_3$} & \multicolumn{1}{l|}{$\beta_4$} & \multicolumn{1}{l|}{$\beta_1$} & \multicolumn{1}{l|}{$\beta_2$} & \multicolumn{1}{l|}{$\beta_3$} & \multicolumn{1}{l|}{$\beta_4$} \bigstrut\\
    \hline
    VDFR  & 0,0291 (0,0079) & 0,0455 (0,0138) & 0,0115 (0,0199) & 0,0229 (0,0165) & 0,0281 (0,0094) & 0,0451 (0,0142) & 0,0106 (0,01) & 0,0208 (0,0189) \bigstrut\\
    \hline
    FF-VDFR & \textbf{0,0093 (0,0033)} & \textbf{0,0269 (0,0145)} & \textbf{0,01 (0,0046)} & \textbf{0,0195 (0,0101)} & \textbf{0,0093 (0,0039)} & \textbf{0,0228 (0,0141)} & \textbf{0,008 (0,0047)} & \textbf{0,017 (0,0077)} \bigstrut\\
    \hline
    \end{tabular}%
    }
\caption{Mean (standard deviation) of 100 measures of AMSE when the domain follows a negative binomial distribution and the response follows a Poisson distribution.}
\label{B-tables}%
\end{table}%

Regarding the RMSE we can see that the FF-VDFR model outperforms all others in all the scenarios shown in the table except in two cases, when the true functional coefficient is $\beta_2(t,T)$ and $\beta_4(t,T)$, the true functional data is noisy and the sample size is 500. Even in these two scenarios the performance of the FF-VDFR and VDFR model is very similar.

In general, it is expected that filtering the noise by making a basis representation of the sampled curves will improve the performance of the estimation when the true functional data is smooth but will be counterproductive when the true functional data is noisy. 

Regarding the AMSE, the FF-VDFR model outperforms the VDFR model in all the scenarios but three, all corresponding to the smallest sample size $N=100$, where both methods perform similarly. Notice that the differences in cases where the FF-VDFR model outperforms the VDFR model can be significant, for example in the scenarios where the true functional coefficient is $\beta_1(t,T)$. 

\begin{figure}[!htb]
  \centering
    \includegraphics[scale=0.7, keepaspectratio=true]{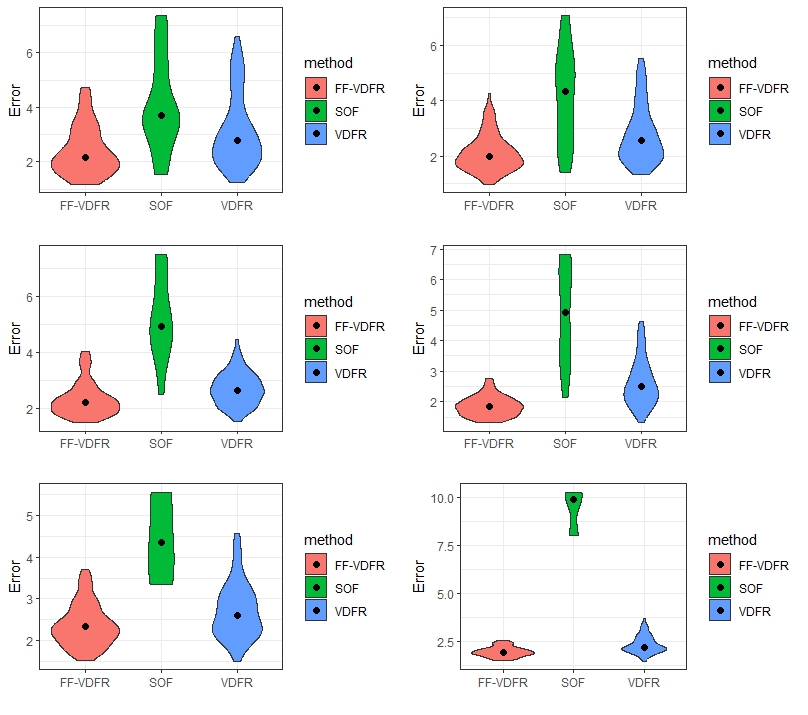}
  \caption{Violin box-plots of the RMSE when the domain follows a negative binomial distribution and the response follows a Poisson distribution and the true functional coefficient is $\beta_3(t,T)$. Left column corresponds with the true functional data being smooth while the right column corresponds with its noisy counterpart. The up, middle, and bottom rows represent sample sizes of $N=100,200,500$, respectively. The dot in the middle of the boxes represents the median value.}
\label{RMSE Beta_3}
\end{figure}

\begin{figure}[!htb]
  \centering
    \includegraphics[scale=0.7, keepaspectratio=true]{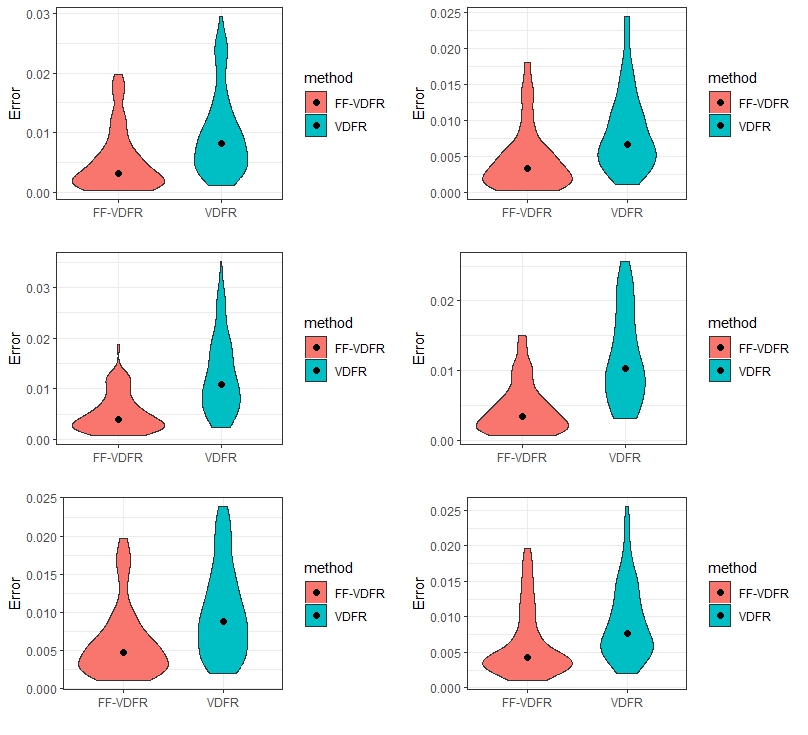}
  \caption{Violin box-plots of the AMSE when the domain follows a negative binomial distribution and the response follows a Poisson distribution and the true functional coefficient is $\beta_3(t,T)$. Left column corresponds with the true functional data being smooth while the right column corresponds with its noisy counterpart. The up, middle, and bottom rows represent sample sizes of $N=100,200,500$, respectively. The dot in the middle of the boxes represents the median value.}
\label{AMSE Beta_3}
\end{figure}

These results could be misleading, because they do not take into consideration the distribution of all the error measures. Figures \ref{RMSE Beta_3} and \ref{AMSE Beta_3} show the violin box-plots for the RMSE and AMSE measures, respectively, for the scenarios when the true functional coefficient is $\beta_3(t,T)$. In these figures, all values that fall outside the interval $(q_1 - 1,5 \cdot s \; ; \; q_3 + 1,5\cdot s)$ have been excluded, with $q_1$ and $q_3$ being the first and third quartile, respectively, and $s$ the standard deviation of the corresponding scenario. 

The results shown in the figures reveal that the FF-VDFR model outperforms all other models for the RMSE and AMSE measurements, in terms of lower error and variability.

From the total of the 96 simulated scenarios, the FF-VDFR model outperformed all the others in 72 scenarios (75\%) in terms of the RMSE. From the scenarios when the proposed methodology was not the best one, 55\% corresponds with noisy true functional data.

Notice that, in the shown scenarios the performance of the FF-VDFR model when is not the best model is only 0.001\% worse than the best. This similar performance when the FF-VDFR model is not the best sustains across the rest of scenarios, being the biggest difference of 0,25\% corresponding to the case when the domain follows an uniform distribution, the true functional coefficient is $\beta_4(t,T)$, the true functional data is noisy, the sample size is 500 and the response follows a Poisson distribution.

Regarding the AMSE, of the total 96 scenarios the FF-VDFR outperformed the VDFR model in 82 scenarios (85,5\%). And from the scenarios where this model did not offer the best performance, 50\% of the cases correspond to noisy true functional data. 

In summary, the FF-VDFR model outperforms all the others in the majority of the scenarios with respect to the two evaluation criteria used. Furthermore, most of the scenarios when the proposed model was not the best in performance correspond with the true functional data being noisy, in which a worse performance of the FF-VDFR model is expected. But even in the 48 scenarios of noisy true functional data, the FF-VDFR model is competitive, outperforming both the VDFR and the SOF models in terms of RMSE in 73\% of the scenarios and outperforming the VDFR models in terms of the AMSE in 85,4\% of the scenarios.    

\section{Case study: The telEPOC dataset}
\label{5}

In this section, we apply the proposed methodology to the telEPOC program set to determine the possible effect of physical activity on the number of hospitalizations due to COPD.

The telEPOC program \citep{Esteban2016OutcomesPatients}
was carried out at the Galdakao-Usansolo University Hospital (Biscay, Spain). Patient collection was done between the years 2010 and 2013, and the study includes five years of follow-up. The main goal of the study was to evaluate the efficacy of a telemonitoring-based program (telEPOC) in COPD patients with frequent hospitalizations. A total of 119 patients defined as those with frequent hospitalizations previous to inclusion were selected for telemonitoring at home.

Moreover, one of the goals of the study was to analyze the effect of performing physical activity on the health of the patients, in particular on the rate of hospitalizations due to COPD. The performance of daily physical activity was measured as the number of daily steps taken by each patient during their time in the study, which was included in the telemonitoring process. This has been the motivation of the work we present in this article.

Patients were included in the study at different time points. However, we are not interested in the effect, if any, that the different dates of admission on patients may have. For that reason, and a clearer analysis, we have aligned to the left all the collected variables making all patients begin the study at “day 1”.

Daily physical activity was analyzed for the 119 patients included initially in the study. The daily steps were recorded only for 112 out of the 119 patients. Two more patients showed too many irregularities in the measurements, therefore they were eliminated. Besides, some of the measurements of the patients were out of the acceptable range of steps that a person can reach during the day, which were considered as missing data. All the missing values for the daily steps were replaced by the mean of the previous and next days. Finally, we worked with a sample of 110 patients with a complete follow-up of daily physical activity. This situation reinforces the idea that observed data present errors and a previous smoothing will provide better results.

The study also collected clinical variables at baseline as possible covariates of interest: smoking habits, age, gender, previous hospitalizations due to COPD, anxiety, and depression symptomatology, among others. The number of hospitalizations due to COPD during the time in the study, the mortality, and the time spent in the study were recorded as potential outcomes. For a detailed explanation of the data collected in the study as well as the enrollment procedures and criteria of acceptance, we refer the readers to \cite{Esteban2016OutcomesPatients}.

The response variable is the number of hospitalizations suffered by each patient and the functional covariate is the daily physical activity for each patient, measured as the daily steps they performed. Then, a fully functional variable domain functional Poisson regression model has been considered. However, the length of the follow-up depends on the patient, and so, the annual rate of hospitalizations was selected as response variable, instead of the number of hospitalizations, in order to avoid the cumulative effect of time in the study. All basis used for this data set are B-splines basis, the number of basis used for the functional covariate was 25 and the number of basis used for the bidimensional coefficient was 625 (25 for each marginal basis).

AIC criterion has been used to select other covariates in the model with adjusting purposes. Nevertheless, for the interpretation of the results, we will focus on the effect of the functional variable on the annual rate of hospitalizations, adjusting by the rest of covariates in the model. The AIC value of the final FF-VDFR model was 187, this value is lower than the AIC value of the VDFR model using the same covariates, which is 309. This result further proves the better performance of the FF-VDFR model over the VDFR model.

When interpreting the functional coefficient a negative value of $\widehat{\beta}(t,T)$ will imply a positive influence on the patient's health, meaning that physical activity is helping to reduce the annual rate of hospitalizations. On the other hand, a positive value would be an indicator of increasing effect on the rate of hospitalizations. However, interpretation should be cautious, without evidence of significant effect in any direction.

\subsection{Results}

The selection criteria end up with four baseline non-functional covariates to be included in the model, namely: gender, previous hospitalizations, anxiety symptomatology and depression symptomatology. Therefore, the final model presents one functional covariate and these four baseline covariates.

In this section we will focus on the results obtained for the estimated functional coefficient $\widehat{\beta}(t,T)$, which reflects directly the relationship between the daily number of steps and the annual rate of hospitalizations due to COPD, adjusted by the baseline covariates in the model.

The estimated functional coefficient is a surface, where the curve resulting of fixing a value of $T = T_i$ represents the influence of physical activity on the annual rate of hospitalizations for patients who have performed this physical activity during $T_i$ days.

\begin{figure}[!htb]
        \centering
		\includegraphics[scale=0.7, keepaspectratio=true]{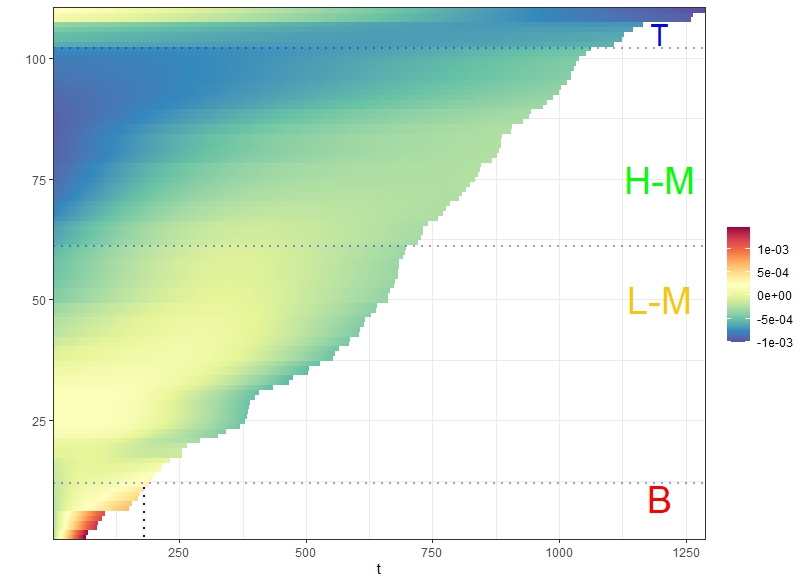}
	\caption{Functional coefficient $\beta(t,T_i) $ for patients with $T_i$ days in the study.}
		\label{Betas}
\end{figure}

Figure 4 shows the heat map of the surface $\widehat{\beta}(t,T)$ for all the periods where patients carried out physical activity. The heat map presents a common feature: doing physical activity regularly during more than 6 months helps to reduce the mean annual rate of hospitalizations due to COPD. This conclusion is based on the fact that all the curves longer than 180 days (the vertical dotted line) end up being negative represented with a cold color (green or blue), meaning a favorable influence of physical activity in the reduction of the annual rate of hospitalizations due to COPD.

For patients whose corresponding curve is shorter than 6 months, represented at the bottom of the heat map and marked with the red letter B, we can see how having performed physical activity this amount of time is not sufficient to reduce the annual rate of hospitalizations since the end of the corresponding curves in the heat map is a warm color (red or orange) meaning a positive value of the curve. 

A more exhaustive examination of the heat map shows several areas where the behavior is similar between patients. The first one already mentioned corresponds with the bottom part of the map; above that can be seen the low-mid area where the patient's influence begins almost null but turns positive towards the end, reflected in the negative values of the curves, this area is marked with the yellow letters L-M. The area directly above is the high-mid region and corresponds to patients whose curves start out negative, reflecting a positive influence on their health, but then increasing their value without ever becoming positive, this area is marked with the green letters H-M. And finally, a small area at the top of the heat map reveals curves that start near zero, i.e., small or null influence over the patient's health, but rapidly decrease meaning that these patients soon begin to notice a positive influence of the physical activity on their health, this region is marked with the blue letter T.

In summary, we may conclude that within this group of patients, performing physical activity helps to reduce the annual rate of hospitalizations due to COPD. More specifically, it was shown that patients who perform physical activity for at least 6 months will see a reduction in their annual rate of hospitalizations.

\section{Discussion}
\label{6}

In this paper a new model to fit variable domain functional data is proposed. This approach is based on assuming the basis representation of both the functional data and the functional coefficient. As a consequence, the functional model turns into a multivariate model, which has been reparametrized to a mixed model to gain computational efficiency. We refer to this new approach as fully functional variable domain functional regression model (FF-VDFR). 

The proposed methodology solves some of the limitations existing in previous approaches such as the optimality of the anisotropic penalties. 

The performance of the FF-VDFR model was tested via a simulation study and compared with the usual scalar-on-function model and the VDFR model, showing that the FF-VDFR model outperforms all the others in the evaluation criteria used.

The methodology presented was developed in order to analyse the influence of physical activity on the annual rate of hospitalizations in COPD patients. The analysis showed that a steady performance of physical activity for at least 6 months helps in the reduction of the annual rate of hospitalizations due to COPD.

It has been demonstrated that there is an association between physical activity, when measured by an accelerometer, and hospitalizations in COPD. In fact, those patients with a high activity (vector magnitude units) had a lower hospitalization risk than those with a low activity (adjusted incidence rate ratio, 0.099; 95\% CI, 0.033-0.293) \citep{GarciaRio2012PrognosticCOPD}. However, as far as we know, there are no studies tracking daily physical activity in COPD and establishing the relationship with hospitalizations. In the telEPOC program daily step count helps clinicians in the management of the patients and in the decision making whenever there is a suspect of exacerbation. Also, this information could support strategies in the mid-term, like boosting physical activity in order to improve health related quality of life of the patients \citep{Esteban2020PredictivePatients}.

A future extension of the FF-VDFR model is the case of the function-on-function regression models for a situation in which the regressors and/or the response variable present variable domain.                 

Our proposal takes into account the variable domain feature in a functional data analysis framework. In particular, it offers an option to analyse functional data with an observation dependent domain, which outperforms other methodologies. As regards to the telEPOC program, clinically relevant results have been obtained, showing that at least six months of continuous physical activity will help to reduce the annual rate of hospitalizations due to COPD, and moreover, to implement interventions in order to keep daily physical activity.

\section{Software}
\label{sec7}

Software in the form of R code and complete documentation is available on request from the corresponding author (pahernan@est-econ.uc3m.es) and can also be found in \url{https://github.com/Pavel-Hernadez-Amaro/V.D.F.R.M-new-estimation-approach.git}. All code use here will be implemented inside an R package.

\section*{Supplementary material}

 Supplementary material is available at
 
\section*{Acknowledgments}

\textit{Conflict of Interest}: None declared.

\section*{Funding}

This work is supported by the grants ID2019-104901RB-I00 and PID2020-113961GB-I00 from the Spanish Ministry of Science, Innovation and Universities MCIN/AEI/10.13039/501100011033. This support is gratefully acknowledged.

\bibliographystyle{biorefs}
\bibliography{FF_VDFR/references}

\end{document}